\documentclass[twocolumn,superscriptaddress,showpacs,showkeys,pra,amsmath,amssymb]{revtex4}

\usepackage{amsfonts,amsmath,amssymb}
\usepackage{graphicx}
\usepackage{hyperref}
\newcommand{\ket}[1]{\left|#1\right\rangle}

\begin{document}
\title{Multi-path entanglement of two photons}
\author{Alessandro Rossi}
\homepage{http://quantumoptics.phys.uniroma1.it/}
\affiliation{
Dipartimento di Fisica, Sapienza Universit\`{a} di Roma,
Roma, 00185 Italy}
\author{Giuseppe Vallone}
\homepage{http://quantumoptics.phys.uniroma1.it/}
\affiliation{
Centro Studi e Ricerche ``Enrico Fermi'', Via Panisperna 89/A, Compendio del Viminale, Roma 00184, Italy}
\affiliation{
Dipartimento di Fisica, Sapienza Universit\`{a} di Roma,
Roma, 00185 Italy}
\affiliation{
Consorzio Nazionale Interuniversitario per le Scienze Fisiche della Materia,
Roma, 00185 Italy}
\author{Andrea Chiuri}
\homepage{http://quantumoptics.phys.uniroma1.it/}
\affiliation{
Dipartimento di Fisica, Sapienza Universit\`{a} di Roma,
Roma, 00185 Italy}
\author{Francesco De Martini}
\homepage{http://quantumoptics.phys.uniroma1.it/}
\affiliation{
Dipartimento di Fisica, Sapienza Universit\`{a} di Roma,
Roma, 00185 Italy}
\affiliation{
Accademia Nazionale dei Lincei, via della Lungara 10, Roma 00165, Italy}
\author{Paolo Mataloni}
\homepage{http://quantumoptics.phys.uniroma1.it/}
\affiliation{
Dipartimento di Fisica, Sapienza Universit\`{a} di Roma,
Roma, 00185 Italy}
\affiliation{
Consorzio Nazionale Interuniversitario per le Scienze Fisiche della Materia,
Roma, 00185 Italy}

\date{\today}
\begin{abstract}
We present a novel optical device based on an integrated system of micro-lenses 
and single mode optical fibers. It allows to collect and direct into many modes two photons
generated by spontaneous parametric down conversion. By this device multiqubit entangled 
states and/or multilevel qu-$d$it states of two photons, encoded in the longitudinal 
momentum degree of freedom, are created. The multi-path photon entanglement realized by 
this device is expected to find important applications in modern quantum information 
technology.
\end{abstract}

\maketitle
 
Entangling two photons in a high-dimension Hilbert space allows the realization of important quantum information tasks. 
These deal with a complete analysis of Bell states 
\cite{schu06prl, barb07pra, wei07pra} and novel protocols of superdense coding \cite{barr08nap}, 
the possibility to perform secure quantum cryptography \cite{bech00prl, thew04qic}
and a fast, high-fidelity one-way quantum computation \cite{brie01prl}, 
\cite{vall07prl, chen07prl, vall08prl, vall08lpl}, besides the realization of 
novel quantum nonlocality tests \cite{coll02prl, yang05prl, barb06prl, kasz00prl}. 

Multidimensional entangled states of two photons have been realized by engineering
both qu-$d$it states and hyperentangled (HE) states. 
In the former each particle belongs to a $d$-level quantum system, in the latter the particles are 
entangled in more than one degree of freedom (DOF). 
Since a qu-$d$it, with $d=2^N$, is equivalent to $N$ qubits, two-entangled qu-$d$its are 
equivalent to the HE state of two particles entangled in $N$ DOFs.
Polarization, time bin and spatial entanglement have been adopted to create qu-$d$its with $d=3$ 
\cite{vall07pra, thew04qic}, $d=4$ \cite{more06prl} and $d=8$ \cite{neve05prl}.
At the same time, HE states have been realized by using in different ways 
polarization, longitudinal momentum and orbital angular momentum, besides time-bin entanglement
\cite{barb05pra, cine05prl, barr05prl, schu06prl, gao08qph}.

\begin{figure*}
\centering\includegraphics[width=17cm]{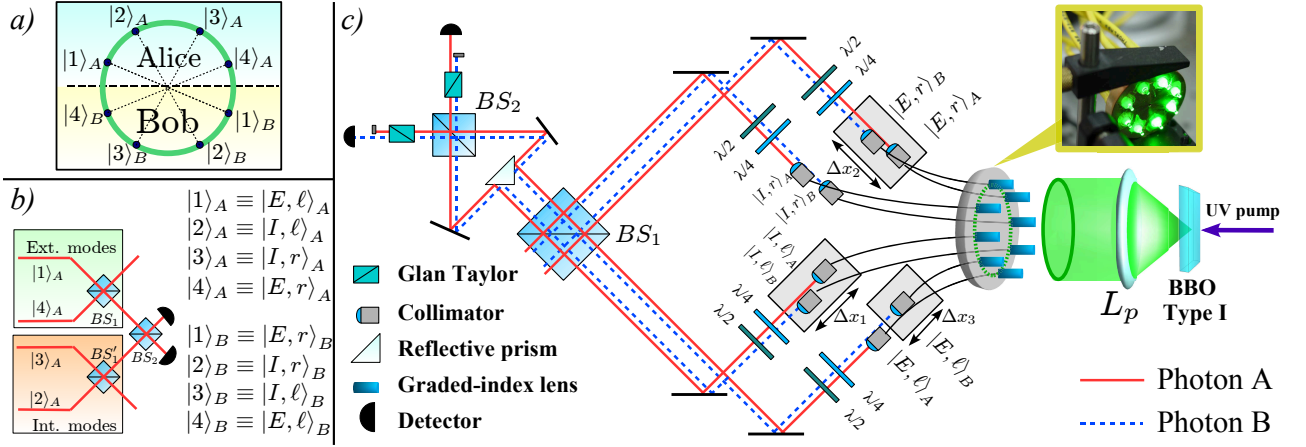}
\caption{$a)$ Labelling of the correlated pairs of SPDC modes.
Photon $A$ ($B$) can be collected with equal probability into one of the four 
modes $\ket{1}_j$, $\ket{2}_j$, $\ket{3}_j$ or $\ket{4}_j$ ($j=A,B$), which 
are relabelled as shown in $b)$.
$b)$ Scheme needed for measurement of photon $A$. 
Here the left ($\ell$) and right ($r$) modes interfere on beam splitters 
$BS_1$ and $BS'_1$, while further interference occurs on $BS_2$ between the 
external ($E$) and internal ($I$) modes. 
An identical scheme is needed for photon $B$.
$c)$ Experimental setup: The GL-SMF integrated 
system injects photon pairs emitted by the crystal into the chained interferometer. 
Path of photon $A$ ($B$) is indicated by the continuous red (dashed blue) line.
By this scheme $BS_1$ plays simultaneously the role of $BS_1$ and $BS'_1$ on side $b)$
for both photons. 
The internal and external modes coming out of the right side of $BS_1$ interfere 
on the beam splitter $BS_2$. Photon $A$ and photon $B$ are separately detected by two 
single photon counting modules.
Waveplates $\lambda/4$ and $\lambda/2$ enable polarization restoration 
of photons after SMF transmission. 
Inset: picture of the 8 GL-SMF system.}
\label{fig:setup}
\end{figure*}

In this paper we present the first experimental realization of a quantum state of
two photons entangled in many optical paths. It is based on the spontaneous parametric 
down conversion (SPDC) emission of a Type I phase-matched nonlinear (NL) crystal 
operating under the excitation of a continuous wave (cw) laser at wavelength (wl) 
$\lambda _{p}$. 
In these conditions, the degenerate signal ($s$) and idler ($i$) photons, are 
generated with uniform probability distribution, at wl $\lambda _{s}=\lambda _{i}=2\lambda _{p}$, 
over a continuum of correlated $\mathbf{k}$-modes belonging to the lateral surface 
of a cone. 
Usually, no more than two correlated spatial modes are used in experiments 
based on Type I crystals, hence the main part of SPDC radiation is lost.
By exploiting the continuum of $\mathbf{k}$-mode emission of Type I crystals, 
high-dimension entangled states can be created \cite{zuko97pra}. 
Indeed, a very large number of qubits are in principle available by this geometry. 
However, a successfull realization of this idea strongly depends on the possibility 
to overcome the practical difficulties represented by independently collecting and 
manipulating the SPDC radiation belonging to a large number of $\mathbf{k}$-modes.

By the device realized in this experiment, photon pairs travelling along a large 
number of $\mathbf{k}$-modes are efficiently coupled into a bundle of single mode optical 
fibers by an integrated system of micro-lenses. 
Four pairs of correlated $\mathbf{k}$-modes have been selected to generate 
two maximally entangled ququads, i.e. qu-$d$its with $d=4$ (or equivalently a $4$-qubit HE state) 
encoded in the longitudinal momentum of the photons:
\begin{equation}\label{mode-hyper}
\ket{\Psi}_{\bf k}=\frac{1}{2}\sum^4_{j=1}
e^{i\phi _{j}}\ket{j}_A\ket{j}_B
\end{equation}
being $\ket j_A$ ($\ket j_B$) the $A$ ($B$) photon mode of the $j$th mode pair and
$\phi_j$ the corresponding phase. Figure \ref{fig:setup}$a)$ shows the annular section of the 
degenerate cone, with four pairs of correlated modes. It is divided in two half-rings, 
respectively corresponding to the Alice ($A$) and Bob ($B$) side. 

This particular geometry allows the creation of multidimensional entangled 
states. However, it is worth to remember that using an increasing number of $\mathbf{k}$-modes necessarily implies 
an exponential requirement of resources since this number doesn't scale linearly 
with the number of qubits. 
In fact, $2^{N}$ $\mathbf{k}$-modes per photon must be selected within the emission
cone to encode $N$ qubits in each photon. 
On the other hand, since a qu-$d$it, with $d=2^N$, is equivalent to $N$
qubits, the number of modes scales linearly with the size $d$ of the state.

We can re-label for convenience the modes $\ket{1}_A,\cdots,\ket{4}_A$ belonging to the $A$ side as
$\ket{E,\ell}_A$, $\ket{I,\ell}_A$, $\ket{I,r}_A$ and $\ket{E,r}_A$, 
where $\ell$ ($r$) and $E$ ($I$) refer to the left (right) and external (internal) 
emission modes (see Figure \ref{fig:setup}$b)$).
They are respectively correlated to the $B$ emission modes 
$\ket{E,r}_B$, $\ket{I,r}_B$, $\ket{I,\ell}_B$ and $\ket{E,\ell}_B$, 
labelled as $\ket{1}_B,\cdots,\ket{4}_B$ in Figure \ref{fig:setup}$a)$.

In the experiment (cf. Figure \ref{fig:setup}c)), a Type I $2mm$ thick $\beta $-barium-borate (BBO) crystal, cut at 
$\theta =51.4$ deg and shined by a vertically polarized cw single longitudinal 
mode MBD-266 Coherent laser ($P=100mW$, $\lambda _{p}=266nm$), produces
degenerate photon pairs ($\lambda _{i}=\lambda _{s}=532nm$), with horizontal 
polarization, along correlated directions belonging to the external 
surface of a cone. 
Then, a positive lens $L_p$ (focal length $f=9.5cm$), located at a distance $f$ from
the BBO, transforms the conical emission into a cylindrical one with 
transverse diameter $D=12mm$.

The device collecting SPDC radiation into a set of single 
mode fibres consists of an integrated system of 8 graded-index lenses (GLs), regularly 
spaced along the circumference of the degenerate ring.
Each GL (Grintech, mod. GT-LFRL-100-025-50-NC (532), length$=2.381mm$, diameter$=1.0mm$, numerical aperture$=0.5$) was glued to a pre-aligned single-mode 
fiber (SMF). 
We measured a coupling efficiency of almost $60\%$ for each GL-SMF system.
Each GL-SMF pair system, corresponding to two correlated $\mathbf{k}$-modes, 
was then glued in a $8$-hole mask, after preliminary 
optimization of its alignment (cf. inset of Figure \ref{fig:setup}c)).
In these conditions, for each GL-SMF pair, nearly $7\cdot10^{3}coinc/s$ 
were measured over a bandwidth $\Delta \lambda =4.5nm$, within a coincidence 
window of $7ns$. This corresponds to a \emph{coincidences/singles} ratio of almost $10\%$.

By this system, the probability that more than one photon pair excites the entire 
mode set within the time coincidence window is related to the number of accidental coincidences, 
hence it grows quadratically with the number of mode pairs, while the real coincidences increase linearly. 
This would contribute to reduce the signal to noise ratio if 
a larger number of mode pairs would be adopted. 

\begin{figure*}
\includegraphics[width=18cm]{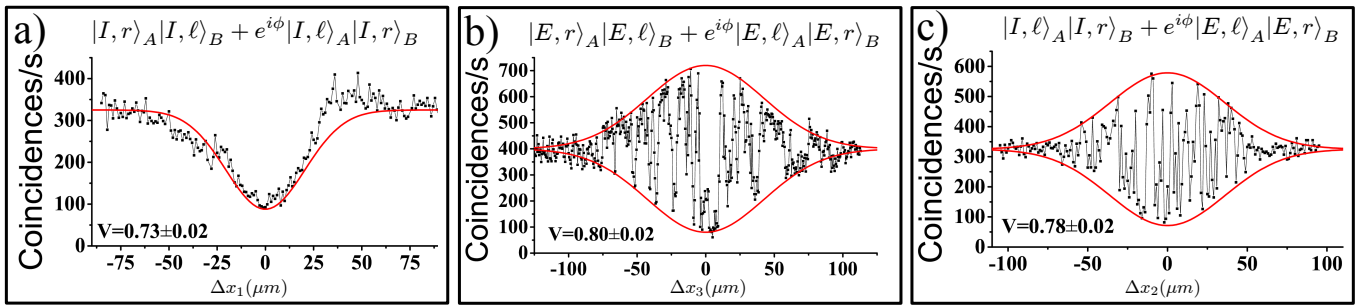}
\caption{Interference patterns obtained by spatial superposition of the internal ($a)$) 
and external ($b)$) modes on $BS_1$ as a function of delays $\Delta x_1$ and $\Delta x_3$.
$c)$: ``Hybrid'' interference condition obtained, varying $\Delta x_2$, by spatial 
superposition on $BS_2$ of internal and external modes.
}
\label{fig:graph}
\end{figure*}
\begin{figure}
\includegraphics[width=8.8cm]{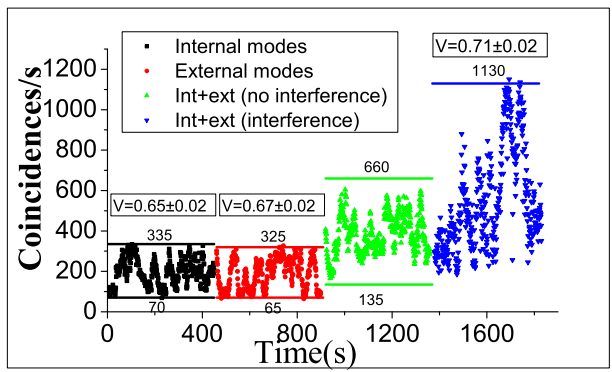}
\caption{Coincidence count oscillations demonstrating multi-path entanglement.
The first two traces correspond to interference occurring on $BS_1$ of internal and external modes,
separately.
The third trace refers to the whole set of mode pairs when no interference occurs on $BS_2$ 
(i.e. for $\Delta x_1=\Delta x_3=0$ and $\Delta x_2\neq0$). 
In the last trace coincidences are measured in conditions of complete mode interference  
($\Delta x_1=\Delta x_2=\Delta x_3=0$). 
The enhancement of visibility, measured by comparing the upper bound of coincidences 
with the maximum that can be obtained in absence of interference, is the signature of the entanglement 
existing between the four pairs of modes.}
\label{fig:interference}
\end{figure}

The entanglement existing between the four mode pairs was characterized by the
interferometric apparatus sketched in Figure \ref{fig:setup}c).
For any couple of mode pairs $\ket{i}_A\ket{i}_B$ and $\ket{j}_A\ket{j}_B$ ($i\neq j$)
we can define a visibility $V^{(ij)}_0=(C_{ii}+C_{jj}-C_{ji}-C_{ij})/(C_{ii}+C_{jj}+C_{ji}+C_{ij})$,
where $C_{ab}$ is the number of coincidences measured between the mode $a$ of photon $A$ and the mode $b$ of photon $B$.
Since almost no coincidence was detected between any pair of non-correlated modes 
(such as, for instance, $\ket1_A$ and $\ket2_B$) we have
$V^{(ij)}_0=0.990\pm0.005$ for any couple of mode pairs. Then, any positive value of visibility measured in any superposition 
basis is sufficient to demonstrate the existence of path-entanglement for any couple of mode pairs.
Similarly, in a 2-qubit entangled state, such as $\frac1{\sqrt2}(\ket{0}_A\ket{0}_B+\ket{1}_A\ket{1}_B)$,
the entanglement witness $W=\openone-\sigma^{(A)}_z\sigma^{(B)}_z-\sigma^{(A)}_x\sigma^{(B)}_x$ \cite{toth05prl} 
reveals the existence of entanglement if the sum of visibilities in the standard basis ($\ket0$ and $\ket 1$) and in the 
superposition basis $\ket{\pm}=\frac{1}{\sqrt2}(\ket0\pm\ket 1)$ is larger than 1.

Indistinguishability between the four modes on which each photon is emitted
can be achieved by the sequence of beam splitters shown in Figure \ref{fig:setup}$b)$.
In the actual chained interferometric setup used in the experiment (cf. Figure \ref{fig:setup}$c)$), 
the $\mathbf{k}$-modes of photons $A$ or $B$, corresponding each to a SMF, 
are matched on the up- and down-side of a common 50:50 beam splitter ($BS_1$), 
where temporal mode matching is realized by fine adjustment of three optical paths by spatial
delays $\Delta x_{1}$, $\Delta x_{2}$, $\Delta x_{3}$. 
Precisely, on the Alice's side, left modes $\ket{I,\ell}_A$ and $\ket{E,\ell}_A$ 
interfere with the corresponding right modes $\ket{I,r}_A$ and $\ket{E,r}_A$. 
The same happens with the corresponding Bob's side modes.
The interference on $BS_1$ derives from spatial and temporal indistinguishability 
obtained by independent adjustment of delays $\Delta x_1$ and $\Delta x_3$, for the internal
and external mode sets, respectively.
This operation acts as a measurement on the $\ell-r$ qubit since the $BS_1$ output modes are 
respectively $\frac1{\sqrt2}(\ket{\ell}_j\pm e^{i\alpha_j}\ket{r}_j)$ ($j=A,B$), 
regardless the value of the $E-I$ qubit.

Interference between internal and external modes is the necessary step to test the the existence of path-entanglement 
over the entire set of $\mathbf{k}$-modes. This was performed by creating interference on the second beam splitter $BS_2$,
i.e. by making indistinguishable the four events 
$\ket{1}_A\ket{1}_B$, $\ket{2}_A\ket{2}_B$, $\ket{3}_A\ket{3}_B$ and $\ket{4}_A\ket{4}_B$. 
Hence, interference can be observed between the output modes of $BS_1$,
i.e. by spatial superposition of the $I$ and $E$ modes on $BS_2$.
Let's consider the input ports of the 50:50 beam splitter $BS_{2}$ in Figure 1$c)$.
There are two interfering modes for the Alice and two for the Bob side, 
coming one from the internal an the other from the external side of the 
first interferometer.
Thus, temporal indistinguishability on $BS_2$ is obtained by varying path delay of the $I$ modes with respect to the $E$ modes. 
This can be achieved by simultaneously varying $\Delta x_2$ and $\Delta x_3$.
Precisely, for any $\Delta x_2=\delta L$ and $\Delta x_3=2\delta L$ (with $\delta L/c$ lower than then pump coherence time), 
we kept at the same time the interference of the external modes on $BS_1$ and varied the delay of external mode with respect to
the internal modes on $BS_2$. 
In this way we were able to simultaneously achieve interference on both $BS_1$ and $BS_2$.
Fine alignment of $BS_{1}$ and $BS_{2}$ allowed accurate phase tuning of the
entangled state. In the experiment, phase stability could be achieved by mechanical stabilization of the
interferometric setup and by thermal heating of the fiber system.
All the measurements presented in this paper were performed without temperature stabilization of the fiber 
apparatus.

The interference patterns given in Figures 2$a)$ and 2$b)$, obtained by respectively varying 
$\Delta x_{1}$ and $\Delta x_{3}$, demonstrate the existence of path-entanglement (on $BS_1$), 
both for the internal and external modes. 
As already explained \cite{barb05pra}, when the delay is simultaneously changed for two modes (as it happens 
by changing $\Delta x_1$ for the internal modes in Figure \ref{fig:graph}$a)$) the state phase is self-stabilized (in this case it is almost equal to
$\pi$). 
This is evident in Figure 2$a)$, where the small asymmetries come from small temperature fluctuations. 
On the other hand, 
by changing only one path, as done by varying $\Delta x_3$ in Figure \ref{fig:graph}$b)$ (or $\Delta x_2$ in Figure \ref{fig:graph}$c)$), 
phase fluctuates randomly.
For the same reason, the FWHM of the interference pattern of Figure 2$b)$ (and 2$c)$) doubles the one 
of Figure 2$a)$.

The two experimental results (with visibilities $V =0.73\pm0.02$ and $V =0.80\pm0.02$) demonstrate 
the entanglement for the internal (2$a)$ and external (2$b)$ modes separately.  The presence of a 
multi-path entanglement over the entire set of modes is demonstrated by the third interference pattern of Figure 2$c)$ ($V =0.80\pm0.02$) occurring on $BS_{2}$ 
between $\ket{I,\ell}_A\ket{I,r}_B$ and $\ket{E,\ell}_A\ket{E,r}_B$, and obtained by varying $\Delta x_{2}$.

As a further demonstration of multi-path entanglement we compare in Figure 3 the coincidence oscillations 
obtained for different values of delays $\Delta x_{1}$, $\Delta x_{2}$ and $\Delta x_{3}$.
In these measurements the phase was almost constant within each data acquisition ($1$ s) but varied 
randomly due to temperature fluctuations from acquisition to acquisition.
The first two oscillations correspond to independent interference of the internal 
($V=0.65\pm0.02$) and external ($V=0.67\pm0.02$) modes on $BS_1$.
The third oscillation corresponds to the coincidences measured in conditions of 
no temporal interference on $BS_2$ (i.e. by setting $\Delta x_1=\Delta x_3=0$ and $\Delta x_2\neq0$).
This must be compared with the fourth oscillation data, 
obtained when $\Delta x_1=\Delta x_2=\Delta x_3=0$, that clearly shows the enhancement of signal due to the 
complete indistinguishability of the four pairs of modes caused by multi-path entanglement. 
The upper bound of coincidence counts ($1130$), compared to the maximum that can be obtained in absence 
of interference ($660$), corresponds to a total visibility $V=0.71\pm0.02$.

In conclusion, we presented in this paper the first experimental realization of a multi-path entanglement of two photons. 
The experiment was performed by collecting four pairs of correlated modes of the degenerate SPDC radiation 
emitted by a NL crystal cut for Type I phase matching.
On this purpose we successfully realized an integrated system of eight graded index lenses coupled to a corresponding 
set of single mode optical fibers.
We tested the entanglement generated over the entire SPDC conical emission of the crystal by using an especially 
developed chained multi-path interferometer.

The novel device presented in this paper can be extended to an even larger number of $\mathbf{k}$-modes by exploiting 
the continuum emission of the crystal and used together with an integrated optical 
circuit to realize miniaturized quantum optical networks of increasing complexity \cite{poli08sci}.
      
Besides multiqubit entanglement, this system can be successfully used in other quantum information 
processes. As an example, it can be adopted in the realization of a ``quasi-deterministic'' 
source of heralded single photons emitted along one half of the SPDC modes and triggered 
by the photons emitted along the other half of correlated modes \cite{migd02pra}.
The same device allows the efficient multiplexing transmission to many pairs of users and over long fiber distances
of time-bin entanglement, which is not affected by thermal or mechanical fiber instabilities \cite{bren99prl}.
 
This device is a useful tool for an efficient generation and distribution of entanglement
since it allows to maximize the emission of photon pairs for a given value of the pump power.
Other configurations can be used on this purpose, for instance those based on a sequence of mirrors and 
beam splitters or several nonlinear crystal, but at the price of a strong reduction of SPDC coincidences.


\end{document}